\documentclass[]{aa}
\usepackage{psfig}
\usepackage{natbib}
\usepackage{epsf}
\usepackage{graphicx}

\bibpunct{(}{)}{;}{a}{}{,}
 
\begin{document}

\def\eps{\varepsilon}
\def\aap{A\&A}
\def\apj{ApJ}
\def\apjl{ApJL}
\def\mnras{MNRAS}
\def\aj{AJ}
\def\nat{Nature}
\def\aaps{A\&A Supp.}
\def\me{m_\e}
\def\lesssim{\mathrel{\hbox{\rlap{\hbox{\lower4pt\hbox{$\sim$}}}\hbox{$<$}}}}
\def\gtrsim{\mathrel{\hbox{\rlap{\hbox{\lower4pt\hbox{$\sim$}}}\hbox{$>$}}}}

\newcommand{\hq}{\hbar}
\newcommand{\epsB}{\varepsilon_{\rm B}}
\newcommand{\epsCMB}{\varepsilon_{\rm cmb}}

\title{Number counts in homogeneous and inhomogeneous dark energy models}
\titlerunning{Number counts in homogeneous and inhomogeneous dark
  energy models}
\author{N. J. Nunes\inst{1}, A. C. da Silva\inst{2,3} and N. Aghanim\inst{2}}
\authorrunning{Nunes, da Silva \& Aghanim}
\institute{School of Physics and Astronomy, 116 Church Street S.E.,
University of Minnesota,
Minneapolis,
Minnesota 55455,
U.S.A.
\and  IAS-CNRS, B\^{a}timent 121, Universit\'{e} Paris Sud, F--91405, Orsay,
France
\and
Centro de Astrof\'{ii}sica da Universidade do Porto, Rua das Estrelas,
4150-764 Porto, Portugal}
\date{\today}

\abstract{In the simple case of a constant equation of state, redshift
distribution of collapsed structures may constrain dark energy models.
Different dark energy models having the
same energy density today but different equations of state give quite
different number counts. Moreover, we show that introducing the
possibility that dark energy collapses with dark matter
(``inhomogeneous'' dark energy) significantly complicates the picture. We
illustrate our results by comparing four dark energy models to the
standard $\Lambda$-model.  We investigate a model with a constant
equation of state equal to $-0.8$, a phantom energy model and two scalar
potentials (built out of a combination of two exponential
terms). Although their equations of state at present are almost
indistinguishable from a $\Lambda$-model, both scalar potentials
undergo quite different evolutions at higher redshifts and give
different number counts. We show that phantom dark energy induces
opposite departures from the $\Lambda$-model as compared with the
other models considered here. Finally, we find that inhomogeneous dark
energy enhances departures from the $\Lambda$-model with maximum
deviations of about 15\% for both number counts and integrated number
counts. Larger departures from the $\Lambda$-model are obtained for
massive structures which are rare objects making it difficult to
statistically distinguish between models.   }

\maketitle

\section{Introduction}\label{sec:intro}

In the present general picture of cosmology, converging evidences
suggest that the matter density parameter is low and that the largest
fraction of the energy density of the universe has an unknown nature
leading to an accelerating phase. These indications come primarily
from supernovae Ia data
(e.g. \cite{Riess:1998cb,Perlmutter:1998np,Riess:2004nr}) and are
corroborated by cosmic microwave background radiation
(e.g. \cite{Spergel:2003cb}) and large scale structure observations
(e.g.\cite{Cole:2005sx}), although there are different interpretations
for the data (e.g. \cite{Blanchard:2003du,Shanks:2004af}).  A
cosmological constant can explain the acceleration of the universe,
however, the disagreement by $\sim 120$ orders of magnitudes with
predictions from theoretical particle physics has shown the need for
resorting to an alternative explanation. This is why several
theoretical models were recently proposed to explain this dark energy
in the universe. This new component can be identified with a slowly
varying, self-interacting, neutral scalar ``quintessence'' field
(\cite{Wetterich:1988fm,Ratra:1988rm,Ferreira:1997au, Zlatev:1998tr})
which can be minimally coupled, non-minimally coupled
(e.g. \cite{Uzan:1999ch,Amendola:1999er,Baccigalupi:2000je}), a
phantom (e.g. \cite{Caldwell:1999ew}), a tachyon
(e.g. \cite{Bagla:2002yn}) or of purely kinetic nature
(e.g. \cite{Armendariz-Picon:2000dh}), known as K-essence.  An
alternative to dark energy is a fluid with a Chaplygin gas type of
equation of state (e.g. \cite{Kamenshchik:2001cp}). See
\cite{Sahni:2004ai} and references therein for a review on some of
these, and other, models.

Regardless of its nature, dark energy as a dominant component, plays a
role in the structure formation and thus is likely to modify the
number of formed structures.  The evolution of linear perturbations in
a scalar field like quintessence and the effects on structure
formation have already been investigated theoretically
(e.g. \cite{Ferreira:1997au,Perrotta:2002sw,Amendola:2003wa}).  The
effects on the abundance of collapsed structures and its evolution
with redshift were also widely explored and suggested as a tool to
constrain the dark energy's nature and evolution (e.g. 
\cite{2001ApJ...553..545H,Weller:2001gk,Weinberg:2002rd,Battye:2003bm,Wang:2004pk,Mohr:2004zh,Horellou:2005qc}).

Recently, numerical simulations including a dark energy component were
performed by several groups to complement the analytical computations
and to study the effects of dark energy at the structure level (e.g. shape
of the dark matter halo, mass function)
(\cite{Linder:2003dr,Lokas:2003cj,Klypin:2003ug,Kuhlen:2004rw}). In
such studies, the scalar field associated with dark energy is assumed
not to have density fluctuations on scales of galaxy clusters or
below. If dark energy influences the perturbations on small scales as
proposed for example by \cite{Arbey:2001qi}, \cite{Bean:2002kx},
\cite{Padmanabhan:2002sh} or \cite{Bagla:2002yn}, the collapse of
structures as well as their properties will be
affected.

\cite{Mota:2004pa} have shown that the properties of collapsed halos
(density contrast, virial radius) depend strongly on the shape of the
potential, the initial conditions, the time evolution of the equation
of state and {\it on the behaviour of the scalar field in non-linear
regions.} This is what we will refer to as the {\it inhomogeneity} of
the scalar field. More recently, \cite{Nunes:2004wn} have investigated
how inhomogeneous quintessence models have a specific signature even
in the linear regime of structure formation. They have shown that the
time of collapse is affected by the inhomogeneity of dark energy and
they have computed the resulting effect on the linearly extrapolated
density threshold $\delta_c$. Moreover, they examined the evolution of
matter overdensity as a function of time varying equation of state in
homogeneous and inhomogeneous assumptions. \cite{Maor:2005hq} have
generalized the formalism to allow for a smooth transition between the
homogeneous and inhomogeneous scenarios. They have concluded
that, if only matter virializes, the final size of the system is
fundamentally distinct from the one reached if the full system
virializes.

In the present study, we extend the work of \cite{Nunes:2004wn} to
investigate how the quintessence field affects the abundance of
collapsed halos when the field follows the background evolution
(homogeneous) and more specifically when it collapses with the dark
matter (inhomogeneous). We compare the two assumptions for models with
constant equation of state and more general cases of time-varying
equation of state. To compute the structure abundances and their
evolution with redshift, we use the canonical \cite{1974ApJ...193..437P}
formalism. Its theoretical expression allows us to account for
the effects of inhomogeneous quintessence field through $\delta_c$,
the growth factor as well as the volume element. Finally, we focus on
the effects of the different models for structures with masses ranging
between $10^{13}$ and $10^{16}\, h^{-1} M_\odot $.

This paper is organised as follows. In Section 2 we introduce the
fundamental equations that describe the evolution of the quintessence
field and the collapse of structure in the homogeneous and
inhomogeneous hypothesis. In Section 3 we describe the method used to
compute the number density of collapsed objects (mass function) in
both of these scenarios, including the case where the equation of
state of dark energy is allowed to vary with time. 
We give results and discuss the effects of the normalisation of the
mass function on the predicted number counts in Section 4, and
present concluding remarks in Section 5.

\section{Theoretical background} \label{sec:th}

In a spatially flat Friedmann-Robertson-Walker Universe the cosmic
dynamics is determined by a background pressureless fluid (dark and
visible matter), radiation and dark energy. The governing equations of
motion are
\begin{eqnarray}
\dot{H} &=&- \frac{\kappa^2}{2} \left( \rho_B^{}+p_B^{} + \rho_{\rm de}
+ p_{\rm de} \right) \,, \\
\label{eqcontinuity}
\dot{\rho}_B^{} &=& -3 H (\rho_B^{}+p_B^{}) \,, \\
\label{rhophieq}
\dot{\rho}_{\rm de} &=& -3H(\rho_{\rm de}+ p_{\rm de}) \,,
\end{eqnarray}
with
\begin{eqnarray}
H^2 = \frac{\kappa^2}{3} \left(\rho_B^{} + \rho_{\rm de} \right) \,.
\end{eqnarray}
%
Here $H = \dot{a}/a$ is the expansion rate of the Universe, $a(t)$ is
the scale factor, $\kappa^2 = 8 \pi G$ and $\rho_B^{}$ and $p_B^{}$
are the energy density and pressure of the background fluid,
respectively. In this work the background is taken to be dominated by
non-relativistic matter, hence, $\rho_B^{}= \rho_{\rm m} \propto a^{-3}$.
If dark energy is a perfect fluid its energy density and pressure are
related by the equation of state $\rho_{\rm de} = w \rho_{\rm de}$ and
$\rho_{\rm de} = \Omega_{\rm de}\rho_0/a^{3(w+1)}$. Alternatively,
dark energy can be described by a dynamical evolving scalar field
rolling down its potential $V(\phi)$. In this case, its energy density
and pressure are defined as, $\rho_{\phi}^{} = \dot{\phi}^2/2 + V$ and
$p_{\phi}^{} = \dot{\phi}^2/2 - V$, respectively where the relation $w
= p_{\phi}/\rho_{\phi}$, still holds.  The equation of motion for the
scalar field is,
\begin{equation}
\label{eqscalar}
\ddot{\phi} = - 3 H \dot{\phi} - \frac{d V}{d \phi} \,.
\end{equation}

We explore models of dark energy that have a significant contribution
at high redshift unlike models such as the inverse power law
(\cite{Zlatev:1998tr}) or SUGRA models (\cite{Brax:1999gp}). We
compare the results of such models with the cosmological constant
model with $w = -1$ and we assume $\Omega_{\rm de} = 0.7$ today for
all models considered in the present study. Moreover, we choose
the present value of the equation of state, $w_0$, and its running, $
dw/dz(z=0)$, to be within the current observational limits
(\cite{Riess:2004nr}).

\begin{figure}
\includegraphics[width=8.8cm]{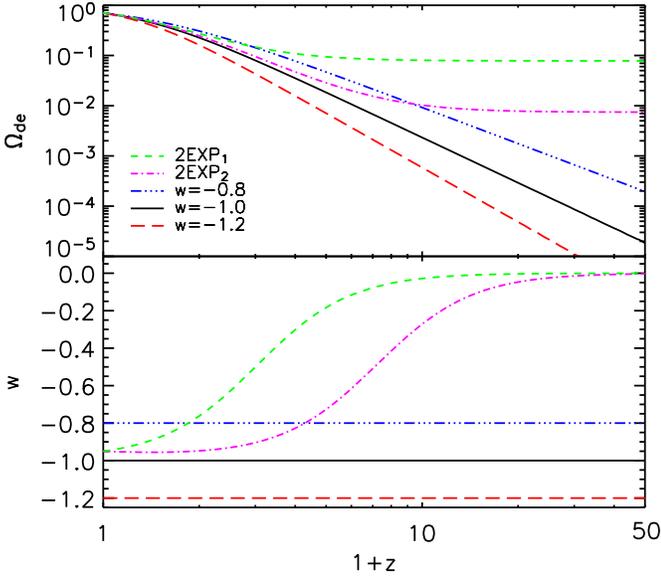}
\caption{Evolution of the dark energy density (top panel) and equation
of state (bottom panel) with redshift for the models
considered in this paper: $w = -1$ (solid line), $w
 = -1.2$ (long dashed line), $w = -0.8$ (triple
dot-dashed line), 2EXP$_1$ $(\alpha,\beta)=(6.2,0.1)$ (dashed
line), and 2EXP$_2$ $(\alpha,\beta)=(20.1,0.5)$ (dot-dashed line).}
\label{fig:omegaw}
\end{figure}

Figure \ref{fig:omegaw} shows the evolution of the dark energy density
and equation of state with redshift for the models we consider in this
paper.
We focus on dark energy models for which $w =
-0.8$, $w = -1.2$ (phantom energy, \cite{Caldwell:1999ew}) and
two cases where the dark energy results from a slowly evolving scalar
field in a potential with two exponential terms (2EXP)
(\cite{Barreiro:1999zs})
\begin{equation}
V(\phi) = V_0 \left( e^{\alpha \kappa \phi} + e^{\beta \kappa \phi}
\right) \,.
\end{equation}
We choose as in Nunes \& Mota (2004) the pairs $(\alpha,\beta) =
(6.2,0.1)$ (2EXP$_1$) and $(\alpha,\beta) = (20.1,0.5)$ (2EXP$_2$).
Both provide an equation of state at present $w_{0} = -0.95$.  The
equation of state for 2EXP$_1$ approaches zero faster than for
2EXP$_2$ (see bottom panel in Fig.~\ref{fig:omegaw}).  The two models
differ also by their contributions at high redshift. As seen from top
panel of Fig.~\ref{fig:omegaw}, 2EXP$_1$ provides a contribution of
dark energy that is non negligible at high redshifts
($\Omega_{\phi}=0.1$ at $z=5$ for 2EXP$_1$ whereas
$\Omega_{\phi}=0.02$ for 2EXP$_2$ at same redshift).

In this work we use the spherical collapse model to describe the
gravitational collapse of an overdense region of radius $r$ and
density contrast $\delta$ such that $1+ \delta = {\rho_{\rm
m}}_c/\rho_{\rm m} = (a/r)^3$, where ${\rho_{\rm m}}_c$ and $\rho_{\rm
m}$ are the energy densities of pressureless matter in the cluster and
in the background, respectively. We have considered the possibility
that dark energy also clusters, i.e. ${\rho_{\rm de}}_c \neq \rho_{\rm
de}$.  Following Mota and van de Bruck (2004) and Nunes and Mota
(2004), we study two extreme limits for the evolution of dark energy
in the overdensity region.  First, we assume that dark energy is
``homogeneous'', i.e. the value of $\rho_{\rm de}$ inside the
overdensity is the same as in the background.  Second, dark energy is
``inhomogeneous'' and collapses with dark matter. In general terms,
the evolution of dark energy inside a cluster can be written as in Mota and
van de Bruck (2004)
\begin{equation}
\label{rhophiceq}
\dot{\rho}_{{\rm de}_{c}} = -3 \frac{\dot{r}}{r}({\rho_{\rm de}}_c+ {p_{\rm de}}_c) +
\Gamma_{\rm de} \,,
\end{equation}
where $\Gamma_{\rm de}$ represents the dark energy loss inside the
cluster and the ratio $\dot{r}/r$ is related to the Hubble ratio and
the evolution of the density contrast through
\begin{equation}
\label{dotdelta}
\frac{\dot{r}}{r} = \frac{\dot{a}}{a} -
\frac{1}{3}~\frac{\dot{\delta}}{1+\delta} \,.
\end{equation}
If dark energy is homogeneous, then
\begin{equation}
\label{Gammaeq}
\Gamma_{\rm de} = -3 \left( \frac{\dot{a}}{a} - \frac{\dot{r}}{r}
 \right) ({\rho_{\rm de}}_c+ {p_{\rm de}}_c) \,.
\end{equation}
However if dark energy is inhomogeneous and collapses with dark
matter, $\Gamma_{\rm de} = 0$ and the equation of motion of a
scalar field inside the cluster is
\begin{equation}
\label{eqscalarc}
\ddot{\phi}_c = - 3 \frac{\dot{r}}{r} \dot{\phi_c} - \frac{d
V(\phi_c)}{d \phi_c} \,.
\end{equation}
Finally, the evolution of the linear density contrast $\delta_L$ is
determined by
\begin{eqnarray}
\label{deltaleq}
\ddot{\delta}_L = - 2 H \dot{\delta}_L
+ \frac{\kappa^2}{2}\left[ \rho_{m} \delta_L + (1+3w)\delta_{\rm de} ~\rho_{\rm de} +3 \rho_{\rm de} \delta w \right]  \,, \nonumber \\
\end{eqnarray}
where we have defined the linear density contrast in dark energy as $\delta_{\rm de} = \delta \rho_{\rm de}/\rho_{\rm de}$ and $\delta w$ is the linear perturbation in the equation of state.

\section{Mass function} \label{sec:massf}

In hierarchical models, cosmic structures form from the gravitational
amplification of small initial density perturbations. The time
evolution of the structure abundances is determined mainly by the rate
at which the perturbations grow until they reach the collapse, or
virialization.  

An analytical computation, proposed by \cite{1974ApJ...193..437P}, gives
the comoving number density of collapsed dark matter halos of mass
$M$ in the interval $d M$ at a given redshift of collapse, $z$ by 
\begin{eqnarray}
\frac{dn}{dM} &=& -\sqrt{\frac{2}{\pi}}\frac{\rho_{\rm m 0}}{M}
\frac{\delta_c(z)}{\sigma (M,z)}
\frac{d\ln\sigma (M,z)}{d M}
\exp\left[-\frac{\delta_c(z)^2}{2\sigma (M,z)^2}\right] \,, \nonumber \\
\label{eq:mf}
\end{eqnarray}
where $\rho_{\rm m0}$ is the present matter mean density of the
universe and $\delta_c(z)$ is the linearly extrapolated density
threshold above which structures collapse, i.e.  $\delta_c(z) =
\delta_L(z = z_{\rm col})$. In an Einstein-de Sitter Universe, an
overdensity region collapses with a linear contrast $\delta_c = 1.686$
(see e.g. \cite{1993sfu..book.....P}). This canonical value was used
in the seminal paper of \cite{1974ApJ...193..437P}. The linearly
extrapolated density threshold has recently been re-computed (Nunes \&
Mota 2004) in a more general case accounting for homogeneous and
inhomogeneous dark energy. They show significant variations of
$\delta_c(z)$ with redshift and with dark energy models considered.
In the present study, we compute halo abundances using the results on
$\delta_c(z)$ obtained by Nunes \& Mota (2004), to which we refer the
reader for further details.

The quantity $\sigma (M,z) = g(z)\sigma_M$ is the linear theory {\it rms}
density fluctuation in spheres of radius $R$ containing the mass
$M$ and $g(z) = \delta_L(z)/\delta_L(z=0)$ is the linear growth
factor.
The smoothing scale
$R$ is often specified by the mass within the volume defined by the
window function at the present time (e.g. Peebles 1980). In our
analysis the variance of the smoothed overdensity containing a mass
$M$ is given by
\begin{equation}
\sigma_M = \sigma_8 \left(\frac{M}{M_8}\right)^{-\gamma/3} \,,
\label{eq:sigma}
\end{equation}
where $M_8 = 6 \times 10^{14} \Omega_{\rm m} h^{-1} M_{\odot}$, the
mass inside a sphere of radius $R_8 = 8 h^{-1} {\rm Mpc}$, and
$\sigma_8$ is the variance of the overdensity field smoothed on a
scale of size $R_8$.  The index $\gamma$ is a function of the mass
scale and the shape parameter, $\Gamma $, of the matter power spectrum
(\cite{Viana:1995yv})
\begin{equation}
\gamma = (0.3 \Gamma + 0.2) \left[ 2.92 + \frac{1}{3} \log
  \left(\frac{M}{M_8}\right) \right] \,.
\label{eq:gamma}
\end{equation}
In our study we use $\Gamma = 0.167$ (Spergel et al. 2003). 

For a fixed $\sigma_8$ (power spectrum normalization) the predicted
number density of dark matter halos given by the above formula
is uniquely affected by the dark energy models
through the ratio $\delta_c(z)/g(z)$. The underlying
assumption of this approach is that the transfer function used in the
computation of $\sigma_M$ is that of a cosmological constant
model. This is a good approximation at cluster scales for
homogeneous dark energy models (\cite{Ma:1999dw}), which remains to
be theoretically investigated in the inhomogeneous hypothesis.

For a fixed local halo abundance, the important quantity to consider
is thus $\xi=\delta_c(z)/\sigma_8 g(z)$.
We have verified that apart from $w=-1.2$ (phantom energy) all
homogeneous models have a ratio $\delta_c/\sigma_8 g$ below that of the
$\Lambda $-model. This means that, at higher redshifts, models $w=-0.8$,
2EXP$_1$ and 2EXP$_2$ are expected to give larger halo densities (whereas
the phantom energy model is expected to give lower abundances) when
compared to the cosmological constant model. For inhomogeneous
dark energy we find that all models, except the $w=-0.8$ model, have
larger $\delta_c/\sigma_8 g$ than the cosmological constant model, and
therefore we expect at high redshift lower hallo densities for this
models (higher abundances in the case of $w=-0.8$) when compared to the
$\Lambda $-model.

To obtain the same halo abundance at $z=0$ we scaled $\sigma_8$
according to
$\sigma_8=\delta_{c}(z=0)\sigma_{8,\Lambda}/\delta_{c,\Lambda}(z=0)$,
where the index `$_\Lambda $' represents the cosmological constant
model, and we fix $\sigma_{8,\Lambda}=0.9$. Table~\ref{tab:sigparams} lists
the values of $\sigma_8$ obtained in this way. Note the much larger
dispersion of $\sigma_8$ between models in the inhomogeneous case
(third column) as compared to homogeneous dark energy (second column)
where models require practically the same $\sigma_8$ to reproduce the
present-day halo abundance of the $\Lambda$-model.

\begin{table}
\caption{Variance of the overdensity field smoothed on
$8 h^{-1} {\rm Mpc}$ used in the normalization of number counts
to the present-day halo abundance in $\Lambda$-model
with $\sigma_{\rm 8}=0.9$ for homogeneous (second column) and
inhomogeneous (third column) dark energy models.}
\label{tab:sigparams}
\begin{tabular}{lcc}
\hline
              & \multicolumn{2}{c}{$\sigma_8$} \\
model         & homogeneous & inhomogeneous \\
\hline
2EXP$_1$      & 0.897       & 0.876\\
2EXP$_2$      & 0.899       & 0.886\\
$w=-0.8$      & 0.898       & 0.857\\
$w=-1.0$      & 0.900       & 0.900\\
$w=-1.2$      & 0.901       & 0.942\\
\hline
\end{tabular}
\end{table}
It is also interesting to see that the 2EXP$_1$ model requires the
lowest and the $w=-1.2$ model the highest $\sigma_8$ normalizations to
reproduce the same local abundance of halos.  This means that
structures form early, at a slow rate, if the universe is dominated by
phantom dark energy. They form late, at a faster rate, in the 2EXP$_1$
model.

\begin{figure}
\includegraphics[width=8.8cm]{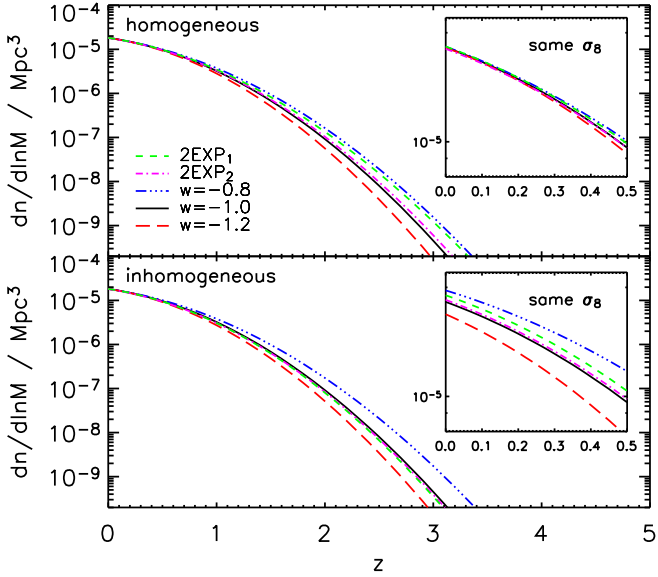}
\caption{Redshift evolution of the modified Press-Schechter mass
function at $M=10^{14} \,h^{-1}\,M_\odot $ for homogeneous (top panel)
and inhomogeneous (bottom panel) dark energy models. In the main
panels models were normalized to reproduce the present-day abundance
of dark mater halos of the $\Lambda-$model with $\sigma_8=0.9$. In the
embedded panels models were normalized to the same
$\sigma_8=0.9$. Lines are the same as for Fig.~\ref{fig:omegaw}.}
\label{fig:mf}
\end{figure}

In Fig.~\ref{fig:mf} we plot the redshift evolution of the mass
function of objects with mass $10^{14}\,h^{-1}\,M_\odot$ for both
homogeneous (top panel) and inhomogeneous (bottom panel) dark energy
using Eq.~(\ref{eq:mf}).
As discussed in the previous paragraphs the $w=-0.8$, 2EXP$_1$ and
2EXP$_2$ models give larger halo abundances than the $w=-1$
model, whereas $w=-1.2$ gives the lowest densities. We
find this same qualitative behaviour within the mass range of interest
for this paper, $10^{13}-10^{16}\,h^{-1}\,M_\odot$, for both homogeneous
and inhomogeneous dark energy models.
The counts were normalized so that at $z=0$ all models give the same
halo abundance as the cosmological constant ($w=-1$) model gives for
$\sigma_8=0.9$. Note that the overdensity contrast at collapse,
$\delta_{c}$, is different for different dark energy models (see
Fig.~2 in Nunes \& Mota 2004) and therefore the number density of
halos at $z=0$ is different if we assume the same $\sigma_8$ for all
models. The embedded panels in Fig.~\ref{fig:mf} show this situation
(zoomed near $z=0$), where $\sigma_8$ was set equal to 0.9 for all
models. Larger differences are found in the case of inhomogeneous dark
energy, because this is where $\delta_{c}(z=0)$ presents larger
deviations from its value in a $\Lambda-$model (see Nunes \& Mota
2004).

In what follows we assume the halo number densities of models
normalized to the present-day halo abundance of the
$\Lambda$-model. In Section~\ref{norm} we discuss the effects of
this assumption on our results.

\section{Predicted number counts}

\begin{figure}
\includegraphics[width=8.8cm]{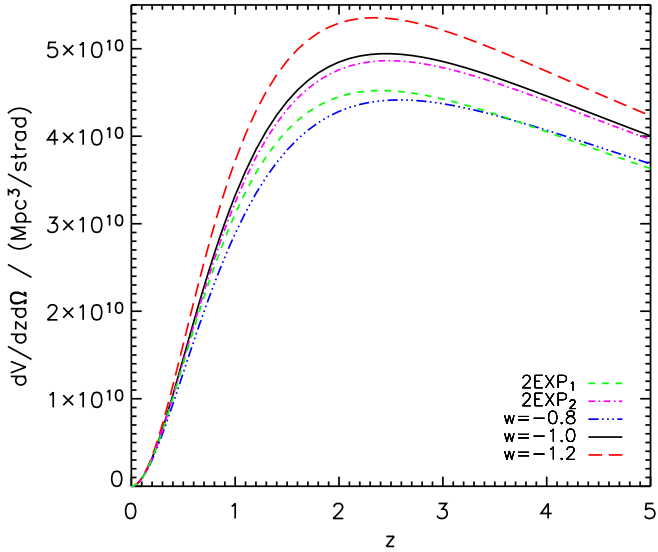}
\caption{Evolution of the volume element with redshift in both
 homogeneous (top panel) and inhomogeneous (bottom panel) dark energy
 scenarios. Lines are the same as for Fig.~\ref{fig:omegaw}. }
\label{fig:dvdzdw}
\end{figure}

In this paper we investigate the modifications caused by a dark energy
component on the number of dark matter halos. We test for one model
with a cosmological constant ($w=-1$) and for four quintessence
models (defined in Section~\ref{sec:th}). The computations are done in
the case where dark energy is homogeneous and in the case where it may
cluster, i.e. inhomogeneous dark energy.
Moreover, we choose to explore the effects on the integrated number of
dark matter halos in mass bins [$M_{\rm inf}, M_{\rm sup}$]
illustrating different classes of cosmological structures, namely
$10^{13}-10^{14}$, $10^{14}-10^{15}$, and $10^{15}-10^{16}$ in units
of $h^{-1}M_\odot $.

We study the effect of dark energy on the number of dark matter halos
by computing two quantities. The first is the all sky number of  halos
per unit of redshift, in the mass bin
\begin{equation}
{\cal N}^{\rm bin}\equiv\frac{dN}{dz} = \int_{4\pi} d\Omega\int_{M_{\rm inf}}^{M_{\rm
                               sup}} \frac{d n}{d M} ~\frac{d V}{d z
                               d\Omega} ~dM \,,
\label{eq:nbin}
\end{equation}
where the volume element is given by $dV/dz d\Omega = r^2(z)/H(z)$,
with $r(z) = \int_0^z H^{-1}(x) dx$. The redshift evolution of $dV/dz
d\Omega $ for different models of dark energy is depicted in
Fig.~\ref{fig:dvdzdw}. Note that the volume element does not depend on the
hypothesis of dark energy clustering, it is thus the same
for both homogeneous and inhomogeneous dark energy models.
Note also that the phantom model is the only one having a volume element
larger than the volume element in the cosmological constant model.

The second quantity we compute is the all sky integrated number counts
above a given mass threshold, $M_{\rm inf}$, and up to redshift $z$:
\begin{eqnarray}
\label{eq:nint}
N(z,M>M_{\rm inf})= \int_{4\pi}~d\Omega\int_{M_{\rm
inf}}^{\infty}\int_0^z \frac{d n}{d M} ~\frac{d V}{d z' d\Omega} ~dM
\,dz'. \nonumber \\
\end{eqnarray}

Our knowledge of both these quantities for galaxy clusters will
improve enormously with upcoming (underway or planned) cluster surveys
operating at different wavebands. 
These include the Planck Surveyor satellite and South Pole Telescope
(SPT) (\cite{Ruhl:2004kv}) Sunyaev-Zel'dovich surveys, the XMM-Newton
serendipitous X-ray cluster survey (XCS) (\cite{2001ApJ...547..594R}) and the
recently proposed deep multiband optical Dark Energy Survey (DES)
designed to probe almost the same sky region of SPT.

\subsection{Number counts in mass bins \label{ncounts}}

\begin{figure}
\includegraphics[width=9cm]{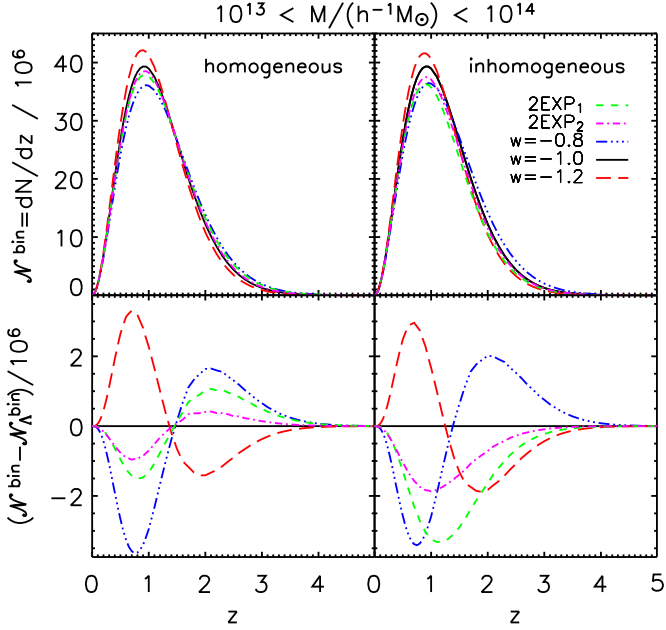}
\caption{ Evolution of number counts (top panels) with redshift and
differences from the $w=-1$ (cosmological constant $\Lambda $)
model (bottom panels) for objects with mass within the range
$10^{13}<M/(h^{-1}M_\odot )<10^{14}$.  Panels on the left show results
for homogeneous dark energy whereas panels on the right show the same
models in the inhomogeneous dark energy scenario.  Lines are the same
as for Fig.~\ref{fig:omegaw}.}\label{fig:nbin-gp}
\end{figure}

\begin{figure}
\includegraphics[width=9cm]{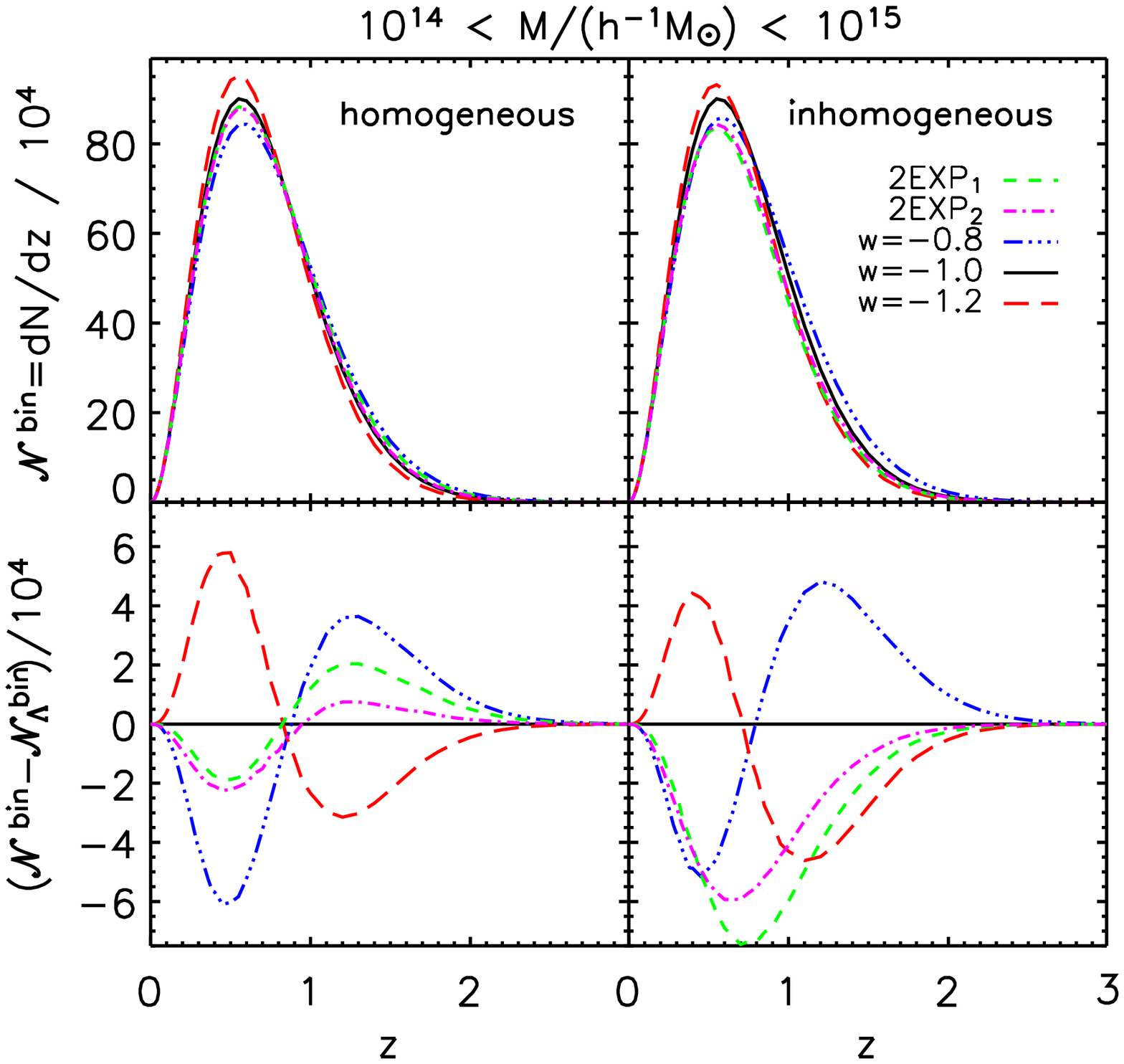}
\caption{ Same as Fig.~\ref{fig:nbin-gp} for objects with mass within
the range $10^{14}<M/(h^{-1}M_\odot )<10^{15}$.  }\label{fig:nbin-cl}
\end{figure}

\begin{figure}
\includegraphics[width=9cm]{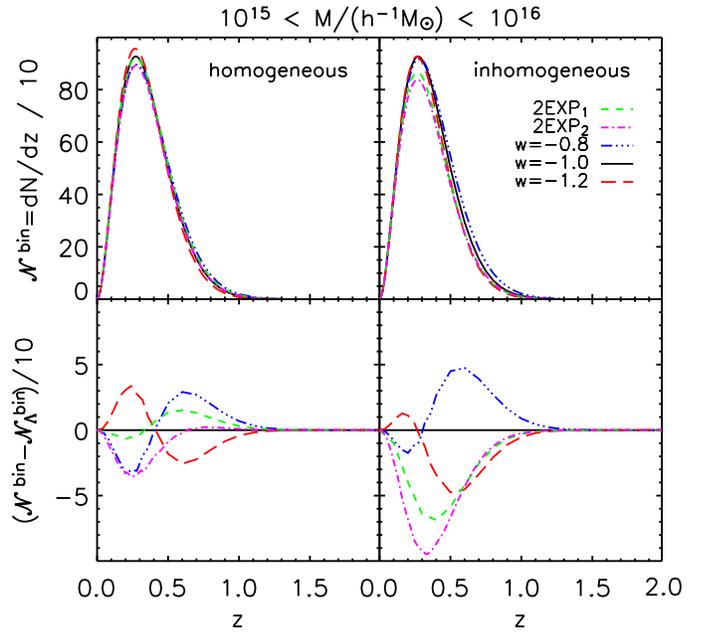}
\caption{ Same as Fig. \ref{fig:nbin-gp} for objects with mass within
the range $10^{15}<M/(h^{-1}M_\odot )<10^{16}$.  }\label{fig:nbin-scl}
\end{figure}

%
Figures \ref{fig:nbin-gp}, \ref{fig:nbin-cl} and \ref{fig:nbin-scl}
show the number counts, ${\cal N}^{\rm bin}=dN/dz$, obtained from
Eq.~(\ref{eq:nbin}) (top panels) together with the difference of
counts of the various dark energy models to the $\Lambda $-model,
$\Delta {\cal N}^{\rm bin}={\cal N}^{\rm bin}-{\cal N}^{\rm
bin}_\Lambda $, (bottom panels).  First let us concentrate on the
homogeneous dark energy case, i.e. on the left panels of these
figures.  Below redshift unity the volume element has the most
important role in the integral of Eq.~(\ref{eq:nbin}), as it increases
by orders of magnitude. Above this redshift the volume element does
not vary much and it is the mass function that decreases by orders of
magnitude. It decreases faster for models of larger
$\delta_c/\sigma_8 g$. Therefore, for our particular models, we expect to
see at low redshifts, in decreasing order of number of counts per
redshift, the following sequence of models: $w =-1.2$, $w=-1$,
$2{\rm EXP}_2$, $2{\rm EXP}_1$ and $w = -0.8$ and the reverse order at
high redshifts. In practice this is only true if the maximum of counts
occurs at a redshift around or greater than unity, i.e after the
volume element does not vary much. We can verify that this requirement
is only satisfied in the lowest mass bin $10^{13} < M/(h^{-1}M_\odot)
<10^{14}$, hence the low redshift order might differ from the one
stated in the two largest mass bins. Nonetheless, the high redshift
description is accurate for all mass bins. More generally, the
differences between counts can be understood in terms of the relative
contribution to ${\cal N}^{\rm bin}$ of the integrand in
Eq.~(\ref{eq:nbin}). The counts thus depend not only on the volume
element and on the growth factor but also on the small differences in
$\sigma_{\rm 8}$.  At high redshift, models with smaller
$\delta_c/\sigma_8 g$ ratios imply larger halo abundances and this effect
dominates that of the volume element.  At low redshifts the
differences between counts from one model to another depend also on
the masses of interest. As a matter of fact, the number of low mass
structures is mostly sensitive to the volume element whereas for
massive structures the number counts become sensitive to the
normalisation $\sigma_{\rm 8}$.  The left lower panels of the
figures illustrate quantitatively the relative evolution of the number
counts. For $w = -1.2$, the figures depict the expected excesses and
deficits of counts compared to the cosmological constant model below
and above redshift $z_{\rm t}$ (defined as ${\cal N}^{\rm bin}(z_{\rm
t}) = {\cal N}_{\Lambda}^{\rm bin}(z_{\rm t})$). Conversely, we have
deficits and excesses below and above $z_{\rm t}$, respectively, for
all the other models. It is also worth pointing out that larger mass bins
imply a larger ratio $\delta_c/\sigma$ (see Eq.~(\ref{eq:sigma})), which makes the
mass function to dominate at lower redshifts in Eq.~(\ref{eq:nbin})
and consequently $z_{\rm t}$ to move to smaller values as depicted in the figures.


\begin{figure}
\includegraphics[width=8.8cm]{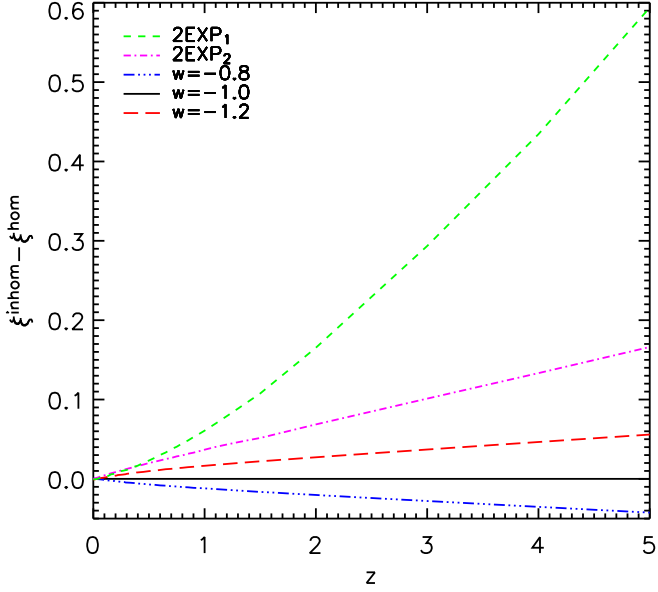}
\caption{Difference of $\xi=\delta_c/g \sigma_8$ between inhomogeneous
($\xi^{\rm inhom}$) and homogeneous ($\xi^{\rm hom}$) dark energy models.}
\label{dgzs8_diff}
\end{figure}

Now, we turn to the comparison between the inhomogeneous and
homogeneous hypotheses (right versus left panels in
Figs.~\ref{fig:nbin-gp}, \ref{fig:nbin-cl} and \ref{fig:nbin-scl}).
Figure \ref{dgzs8_diff} is helpful to understand how the differences
arise. Indeed, when compared to the homogeneous case, we verify that
the quantity $\delta_c/\sigma_8 g$ is smaller in the inhomogeneous case
for $w = -0.8$ and larger for all the other models. 
Given this, it is clear that for $w = -0.8$ we see, as
expected, smaller deficits and larger excesses of counts comparatively
to the homogeneous case and for all the other models the inverse
trend.  The effects of the inhomogeneous hypothesis are more evident
for high mass bins where count deficits and excesses show larger
differences between homogeneous and inhomogeneous dark energy.  For
example, in Fig.~\ref{fig:nbin-scl} we see that the $w=-0.8$ model has
$\sim 2$ times larger count excess in the inhomogeneous case.

%

\subsection{Integrated number counts}
\begin{figure}
\includegraphics[width=9cm]{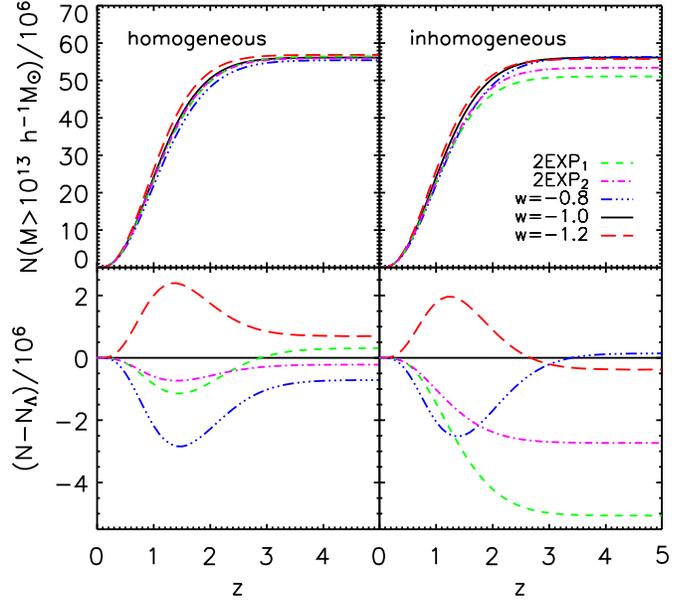}
\caption{\label{Ncounts} Integrated number counts up to redshift $z$
(top panels) and count differences to the $w=-1$ ($\Lambda $) model
(bottom panels) for objects with mass $M>10^{13}\, h^{-1}M_\odot$.
Panels on the left show results for homogeneous dark energy whereas
panels on the right show the same models in the inhomogeneous dark
energy case. Lines are the same as for Fig.~\ref{fig:omegaw}.
}\label{fig:nint-gp}
\end{figure}

\begin{figure}
\includegraphics[width=9cm]{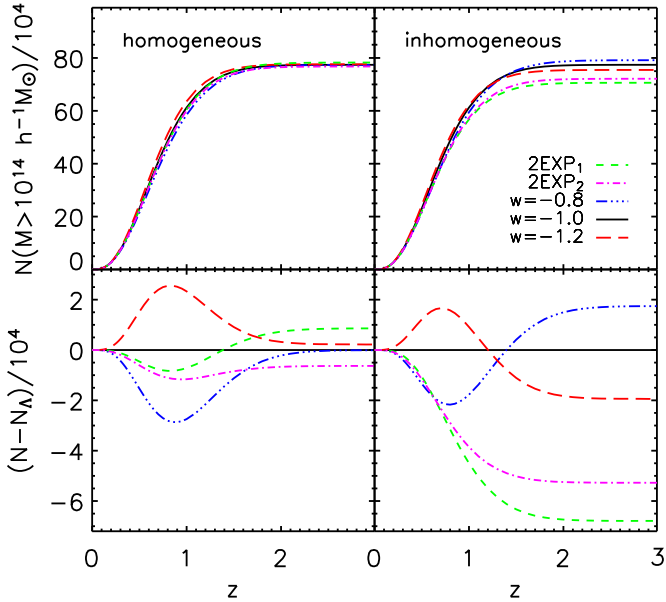}
\caption{ Same as Fig.~\ref{fig:nint-gp} for objects with mass
$M>10^{14}\, h^{-1}M_\odot $. }\label{fig:nint-cl}
\end{figure}

The integrated number of collapsed structures above a given mass
(Eq.~(\ref{eq:nint})) is an important observable quantity. In this
section we present results for the integrated number counts of
structures with masses above $M_{\rm inf}=10^{13}h^{-1}M_\odot$ and
$10^{14}h^{-1}M_\odot$. These are displayed in the upper panels of
Figs. \ref{fig:nint-gp} and \ref{fig:nint-cl}, respectively 
(here we
omit displaying results for $M>M_{\rm inf}=10^{15}h^{-1}M_\odot$
because, as it will be clear below, integrated counts in this mass
range can be directly estimated from the curves in
Fig.~\ref{fig:nbin-scl}).
We also calculate the difference in integrated number counts with
respect to the cosmological constant model shown in the lower panels
of the same figures. As in the previous section, we compare the
predicted numbers in the homogeneous (left panels of figures) and
inhomogeneous hypotheses (right panels) for the cosmological models we
consider in this study.

For each mass bin, the differences between integrated counts 
result from the
combination of effects (that act on different mass ranges) discussed
in the previous section. In our work we have obviously not performed
the integration in the mass range in Eq.~(\ref{eq:nint}) all the way
up to infinity but only to $M_{\rm sup} = 10^{16} h^{-1}M_\odot$.  In
the cases $M>M_{\rm inf}=10^{13}h^{-1}M_\odot$ and $M>M_{\rm
inf}=10^{14}h^{-1}M_\odot$, integrated number counts reflect mainly
the behaviour of curves in the lowest and middle mass bin,
respectively.  This is because $N(z,M>M_{\rm inf})$ is dominated by
the contribution of the lower bound of the mass integration range.
Hence, it is legitimate to concentrate on the lowest mass when a
qualitative description of the integrated number counts is concerned.

We have seen in the Figs. \ref{fig:nbin-gp}, \ref{fig:nbin-cl} and
\ref{fig:nbin-scl} that in the homogeneous case, the model with
$w=-1.2$ presents excesses with respect to the cosmological constant
at low redshift and deficits at high redshift. All the other models
give the opposite behaviour as their $\delta_c/\sigma_8 g$ is always
lower than that of the $w=-1$ model. Therefore, for $w = -1.2$ we
expect the integrated number counts to be larger than for the
cosmological constant model until the redshift $z_{\rm t}$.  The
remaining models must show the opposite behaviour.  The difference
between the integrated number counts of a model with respect to the
cosmological constant must decrease with redshift for $z > z_{\rm t}$.
Eventually above a redshift $z_{\rm flat}$, the integrated number
counts should become constant with redshift because, as we have noted
before, the number counts (Eq. (\ref{eq:nint})) decrease exponentially
with redshift hence contributions above $z_{\rm flat}$ become
negligible. Note that $z_{\rm flat}$ becomes progressively smaller for
structures with larger mass limits ($M_{\rm inf}$) because a smaller
number of these objects form at higher redshifts. This simply reflects
the hierarchical nature of structure formation in models described by
Eqs.~(\ref{eq:mf})-(\ref{eq:gamma}).  Models give quite different
integrated count differences ($N-N_\Lambda $) depending on the maximum
redshift of integration. For homogeneous dark energy, maximum
deviations from the $\Lambda $-model are generally obtained near
$z_{\rm t}$ of the dominant class of structures, whereas for
inhomogeneous dark energy maximum deviations generally occur at much
higher redshifts, $z\simeq z_{\rm flat}$.


In the inhomogeneous scenario, the interpretation of the redshift
dependence of the integrated number counts is similar to that of the
homogeneous case. As we have seen, the inhomogeneous case yields 
higher
excesses and lower deficits for the $w = -0.8$ model, hence we expect
at high redshift, a positive difference in integrated counts
compared to the cosmological constant model. This is quite visible in
the right lower panels of Figs. \ref{fig:nint-gp} and
\ref{fig:nint-cl}. Through a similar argument, we expect at high
redshift, a negative difference of integrated number counts with
respect to the cosmological constant for the $w = -1.2$
model. Moreover, because there are hardly any excesses in the 2EXP
models, we must expect the difference of integrated number counts to
flatten out at $z_{\rm t}$ at large negative values for these two models.

\subsection{Normalisation}
\label{norm}
So far in this work we have normalised the various dark energy models
such that they reproduce the abundance of dark matter halos at
redshift zero (i.e. the local abundance) of a $\Lambda$-model with
$\sigma_8 = 0.9$.  However, observations that are being used to
quantify the number of observed structures trace baryonic (emitting)
gas and not dark matter halos directly. For systems where
non-gravitational physics is important, i.e. the less massive
structures, this may lead to an important miss-match between dark
matter halos and emitting structure counts.  Therefore, we dedicate
the rest of this section to discuss the implications on our results of
dropping the constraint of normalizing models to the same abundance at
$z=0$. Instead, we consider that all models have the same
$\sigma_8=0.9$ normalization, as given by present day observations
(e.g. Spergel et al. 2003).

\begin{figure}
\includegraphics[width=9cm]{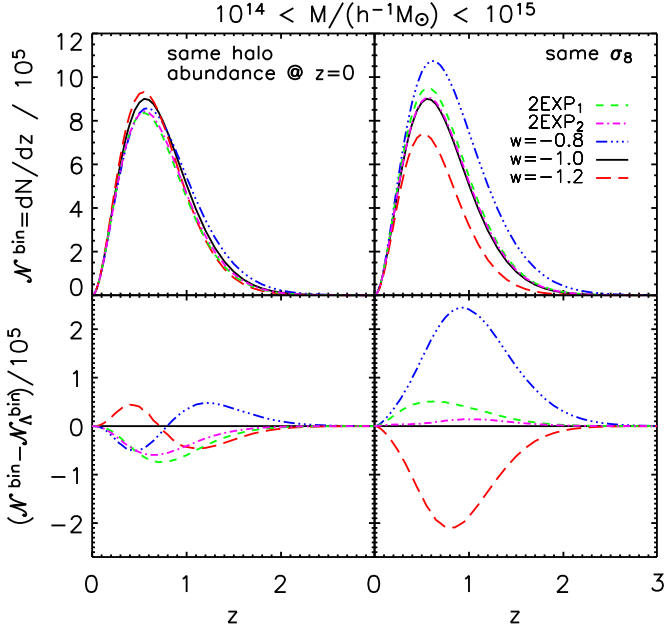}
\caption{Redshift dependence of number counts (top panels) and count
differences relative to the $\Lambda$-model (bottom panels) for objects
with mass within the $10^{14}< M/(h^{-1}M_\odot)<10^{15}$, assuming
dark energy is inhomogeneous. Panels on the left show results assuming
models are normalized to the same halo abundance at $z=0$, whereas
panel on the right assume that models have all $\sigma_8=0.9$ (which
implies different halo abundances at $z=0$, see text). Lines are the
same as in Fig.~\ref{fig:omegaw}. }\label{fig:clusters_dn_comp}
\end{figure}

\begin{figure}
\includegraphics[width=9cm]{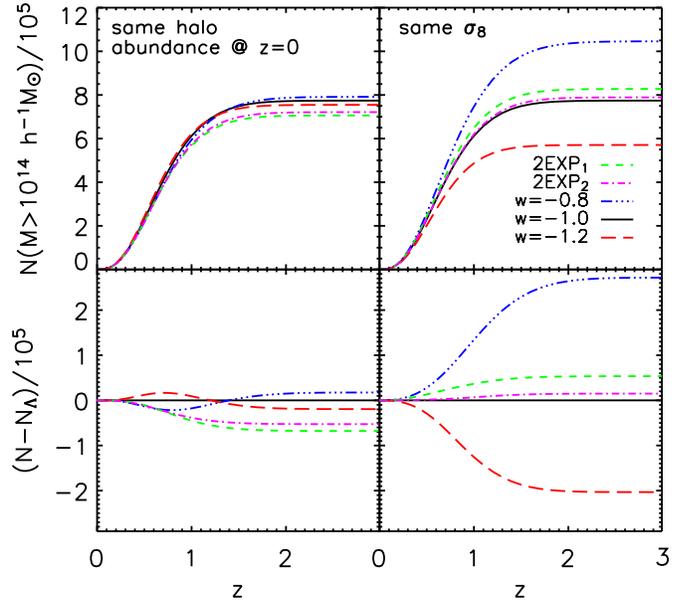}
\caption{Integrated number counts up to redshift $z$ (top panels) and
integrated count differences relative to the $\Lambda$-model (bottom
panels) for objects with mass above $10^{14} h^{-1}M_\odot$, assuming
dark energy is inhomogeneous. Panels on the left show results assuming
models are normalized to the same halo abundance at $z=0$, whereas
panels on the right assume that models have all $\sigma_8=0.9$. Lines
are the same as in Fig.~\ref{fig:omegaw}.}
\label{fig:nclusters_comp}
\end{figure}

We illustrate the effects of inhomogeneous dark energy on one single
mass bin.  Figure \ref{fig:clusters_dn_comp} shows number counts in
the mass bin $10^{14}< M/(h^{-1} M_\odot) < 10^{15}$ and
Fig. \ref{fig:nclusters_comp} the integrated number counts above mass
$M > 10^{14} h^{-1} M_{\odot}$. Both figures consider inhomogeneous
dark energy models.  In the left panels of these figures we are
assuming that models are normalized to the same halo abundance at
$z=0$ (i.e. same curves as those in the right panels of
Figs.~\ref{fig:nbin-cl} and \ref{fig:nint-cl}), whereas right panels
assume that all models have a fixed normalisation, $\sigma_8=0.9$.
Models in the homogeneous hypothesis have practically the same
$\sigma_8$ when they are normalized to the halo abundance at $z=0$.
Therefore we do not expect to observe much difference in the halo
abundances from one dark energy model to another. The situation is
quite different when dark energy is inhomogeneous.  The comparison
between panels in Fig.~\ref{fig:clusters_dn_comp} indicates that
fixing $\sigma_8$ in all models causes much larger departures from the
$\Lambda$-model than in the case where models are normalized to
reproduce the same local halo abundances. 
At the maximum of ${\cal N}^{\rm bin}$, the differences between dark
energy models come from the fact that for a fixed $\sigma_8$, the mass
function reflects the variations of $\delta_c/g$ with $z$, which
effects dominate those of the volume element in Eq.~(\ref{eq:nbin})
for this mass bin. It is interesting to note that the structure of the curves can change dramatically depending weather we fix the local abundance or $\sigma_8$.

As this work was being completed \cite{Manera:2005ct} made public an
analysis similar to the one presented here for the particular scenario
of coupled quintessence. In their work, they have fixed $\sigma_8$ for
the two models under study rather than the local halo abundance. We
have verified that indeed, also in the coupled quintessence model, the
departures from a cosmological constant model are larger if one takes
the former approach.

\section{Conclusions}
Number counts of collapsed structures are commonly proposed as a
tool to probe dark energy models. In the simplest case of a constant
equation of state, galaxy cluster number counts may constrain the dark
energy, provided we have a good knowledge of the cluster physics and
their redshifts (see for example \cite{Majumdar:2003mw,Wang:2004pk}).
In the present study we have
investigated, using the Press-Shechter mass function, two complications
to the general picture: First, we have revisited the modifications
introduced by a varying equation of state or by including a
phantom component; second we have explored the effect of an inhomogeneous
dark energy which collapses with the dark matter.

More specifically, we have considered a scalar potential built out of
two exponential terms for which two sets of parameters (models
2EXP$_1$ and 2EXP$_2$) were explored.  Although their equations of
state at present are almost indistinguishable,
they undergo quite different evolutions at higher redshifts and
generally give different results. In our analysis we have also
included a phantom dark energy model ($w=-1.2$), compatible with
present observations.  An interesting feature of phantom dark energy
is that it induces opposite departures from the $\Lambda$-model as
compared with the other models considered in this paper. That is, we
expect an excess of sources in a phantom energy model when other
models predict a deficit.

Recently Mota and van de Bruck (2004) proposed that dark energy may
cluster in forming structures (inhomogeneous dark energy). They have
investigated the growth and collapse of cosmological structures under
the inhomogeneous hypothesis. Going a step
forward, we have investigated in the present study the implications of
this hypothesis on the number density of collapsed objects.  
We have found that inhomogeneous dark energy generally enhances departures
from the $\Lambda$-model. This includes the models with a time varying
equation of state, which can present several times
larger departures (from the $\Lambda$-model) as compared to the
homogeneous case. Yet our results indicate that the inhomogeneous
dark energy hypothesis causes maximum deviations no larger than $\sim
15$\% in mass bins with comfortably large numbers of collapsed halos.
Another interesting feature is that
maximum departures from the $\Lambda$-model are generally obtained at higher
redshift for inhomogeneous dark energy than for the homogeneous case,
which generally show maximum departures near the maximum of ${\cal
N}^{\rm bin}(z)$. This may be a helpful feature to test for the
inhomogeneous hypothesis.
Larger departures from the $\Lambda$-model are also stronger for the
more massive structures, but these are quite rare objects, which makes
it difficult to statistically distinguish between models.
Our analysis reveals that the inhomogeneous dark energy hypothesis has
the greatest impact on the 2EXP1 model.

In this work, we have assumed that the matter transfer function
remains unchanged at cluster scales. We have further assumed that
models are normalized to reproduce the same abundance of dark mater
halos at redshift zero. In the homogeneous hypothesis all models have
practically the same $\sigma_8$ and there are not much differences in
the halo abundances from one dark energy model to another. When dark
energy is inhomogeneous, $\sigma_8$ differ by a few percent and the
departures from the $\Lambda$-model are much larger. It is, however,
worth noting that the gas physics which rules the observed quantities
adds a degree of degeneracy. We have evaluated the effects of
alternatively fixing $\sigma_8$ to a specific value regardless of the
dark energy model. In this case we verified that the departures from a
$\Lambda$ cosmology are further enhanced.

Our results show that constraining dark energy models from structure
counts is complicated when models have time varying equation of
state. It becomes an even more complicated task when the possibility
of inhomogeneous dark energy is taken into account. Therefore in order
to constrain dark energy models, we need to explore as many observable
quantities as possible. Our results suggest that besides redshift
distribution of structures, considering structures in mass ranges
significantly increases the number of observables.  Indeed, each
theoretical model provides specific predictions for the redshift
evolution of number counts and integrated count differences in
different mass bins. The comparison of such quantities with
observations can be used for testing models against observational
data. It may further allow to distinguish between homogeneous and
inhomogeneous dark energy models. 
However, this requires good
knowledge of the gas physics, redshifts of observed structures and,
more precisely, a good understanding of the selection function of the
observations.

\begin{acknowledgements}
We thank David Mota and Morgan LeDelliou for invaluable discussions.
NJN is supported by the Department of Energy under contract
DE-FG02-94ER40823 at the University of Minnesota. AdS acknowledges
support by CMBnet EU TMR network and Funda\c{c}\~{a}o para a
Ci\^{e}ncia e Tecnologia under contract SFHR/BPD/20583/2004, Portugal.
This work was partly supported by CNES. We would like to thank the
referee for his/her comments.
\end{acknowledgements}

\end{document}